\documentclass[10pt, conference, compsocconf]{IEEEtran}

\usepackage{amssymb}
\usepackage{graphicx}
\usepackage{booktabs} 
\usepackage{url}
\usepackage{flushend}

\begin{document}

\title{Challenges and Recommendations for Preparing HPC Applications for Exascale}

\author{\IEEEauthorblockN{
Erika \'Abrah\'am\IEEEauthorrefmark{1},
Costas Bekas\IEEEauthorrefmark{2},
Ivona Brandic\IEEEauthorrefmark{3}, 
Samir Genaim\IEEEauthorrefmark{4},
Einar~Broch~Johnsen\IEEEauthorrefmark{5},\\
Ivan Kondov\IEEEauthorrefmark{6},
Sabri Pllana\IEEEauthorrefmark{7} and
Achim Streit\IEEEauthorrefmark{6}
}
\IEEEauthorblockA{\IEEEauthorrefmark{1} 
RWTH Aachen University - Aachen, Germany
}
\IEEEauthorblockA{\IEEEauthorrefmark{2} 
IBM Research - Zurich, Switzerland
}
\IEEEauthorblockA{\IEEEauthorrefmark{3} 
Vienna University of Technology, Vienna, Austria
}
\IEEEauthorblockA{\IEEEauthorrefmark{4} 
Complutense University of Madrid, Madrid, Spain
}
\IEEEauthorblockA{\IEEEauthorrefmark{5} 
University of Oslo, Department of Informatics, Norway
}
\IEEEauthorblockA{\IEEEauthorrefmark{6} 
Karlsruhe Institute of Technology, Steinbuch Centre for Computing, Germany 
}
\IEEEauthorblockA{\IEEEauthorrefmark{7} 
Linnaeus University, Department of Computer Science, V\"axj\"o, Sweden 
}
}

\maketitle

\begin{abstract}
While the HPC community is working towards the development of the first Exaflop computer (expected around 2020), after reaching the Petaflop milestone in 2008 still only few HPC applications are able to fully exploit the capabilities of Petaflop systems. In this paper we argue that efforts for preparing HPC applications for Exascale should start before such systems become available. We identify challenges that need to be addressed and recommend solutions in key areas of interest, including formal modeling, static analysis and optimization, runtime analysis and optimization, and autonomic computing. Furthermore, we outline a conceptual framework for porting HPC applications to future Exascale computing systems and propose steps for its implementation. 

\end{abstract}

\section{Introduction}
\label{sec:introduction}

Exascale computing~\cite{transition2exascale} is expected to revolutionize computational science and engineering by providing 1000x the capabilities of currently available computing systems, while having a similar power footprint. The total performance of the 500 systems in the 44th TOP500 list (18 Nov 2014, \url{http://top500.org/}) is about 0.3 exa\-FLOPS. The HPC community~\cite{exascale-software-roadmap} is now working towards the development of the first Exaflop computer, expected around 2020, after reaching the Petaflop milestone in 2008.  However, only a few HPC applications are so far able to fully exploit the capabilities of Petaflop systems~\cite{petascale}. Examples of typical scalability for commonly used HPC applications in our organizations are provided in Table \ref{tab:hpcapps}. As the existing HPC applications are the major HPC asset, it is important and challenging to increase their scalability and lifetime by making them Exascale-ready before 2020.

\begin{table*}[tb]
\centering
\caption{Typical current scalability (in processor cores) of commonly used HPC applications in our organizations.}
\small
\begin{tabular}{l@{\quad}l@{\quad}l@{\qquad}r}
\toprule
\textbf{Code} &  \textbf{Application Domain} & \textbf{Language} & \textbf{Scalability} \\
\midrule
WIEN2k & Materials Science & F90 & 1024 \\ 

SIMONA & Nano Science & C++ & 16384  \\ 

ECHAM/MESSy  & Environmental Science & F77/F90 & 1000 \\ 

CORSIKA & Astroparticle Physics & F77/F90 & 2500 \\ 

OpenFOAM & Computational fluid dynamics & C++ & 16384 \\ 

IBM Watson & Graph Analytics & C++ & 32768 \\ 

Bifrost & Stellar\newline atmosphere simulation & F90 & 6500 \\ 
\bottomrule
\end{tabular}
\label{tab:hpcapps}
\end{table*}

The major challenge for preparing HPC applications for Exascale is that there is no Exascale system available yet. Currently all we have are assumptions about Exascale systems. Therefore the commonly used measurement-based approaches for reasoning about performance issues are not applicable. Pre-exascale systems (known as \emph{Summit} and \emph{Sierra}) that IBM~\cite{ibm_press} is developing for the U.S. Department of Energy will exceed 100 petaflops and may provide hints about the extreme-scale architectures of the future. 

This paper argues that efforts for preparing HPC applications for Exascale should start before such systems become available. We identify challenges that need to be addressed and recommend solutions in areas that are relevant for porting HPC application to future Exascale computing systems, including formal modeling, static analysis and optimization, runtime analysis and optimization, and autonomic computing. 

We suggest that porting of HPC applications should be made by successive, stepwise improvements based on the currently available assumptions and data about Exascale systems. This approach should support application improvement each time new information about future Exascale systems becomes available, including the time when the application is actually deployed and runs on a concrete Exascale system. A high-level application representation that captures key functional and non-functional properties in conjunction with the abstract machine model will enable programmers and tools to reason about and perform application improvements, and will serve as input to runtime systems to handle performance and energy optimizations and self-aware fault management. A tunable abstract machine model encapsulates current assumptions for future Exascale systems and enables a priori application improvement before the concrete execution platform is known. At runtime, the model is a posteriori tuned to support activities such as feedback-oriented code improvement or dynamic optimization. 
 
Major contributions of this paper include,
\begin{itemize}
	\item identification of challenges and recommendation of solutions in formal modeling (Section~\ref{subsec:modeling}), static analysis and optimization (Section~\ref{subsec:static}), runtime analysis and optimization (Section~\ref{subsec:runtime}), autonomic computing (Section~\ref{subsec:autonomic});
		\item a conceptual framework for preparing HPC applications for Exascale that supports \emph{a priori} application improvements before the concrete execution platform is known as well as \emph{a posteriori} optimization at runtime (Section~\ref{sec:approach});
	\item a discussion of the related work (Section~\ref{sec:related-work}). 
\end{itemize}

\section{Challenges and Recommendations}
\label{sec:cr}

In this section we identify challenges and recommend solutions in formal modeling, static analysis and optimization, runtime analysis and optimization, and autonomic computing.

\subsection{Formal modeling}
\label{subsec:modeling}

Our goal is to adapt HPC application code to Exascale execution
platforms to achieve good utilization of resources. For this, we need
to address questions such as: 
\begin{enumerate}
	\item What would happen if we change application or hardware layout?
	\item What would happen if we change some parameters of the execution platform?
	\item What would happen if we use a different execution platform?
\end{enumerate}
Unfortunately, answering these questions cannot be done experimentally
at the concrete level because such platforms do not yet exist. 
An alternative is to address these questions at an abstract level,
focusing only on \emph{relevant information} without actually executing
the program.

We believe that relevant information in this context is not what
the code aims to achieve (the result of the computation) but its
corresponding \emph{resource footprints}, that is, how computational
tasks communicate and synchronize, the amount of resources (such as memory
and computing time) these tasks require, and how they access and move data.

In order to adapt the HPC code to a particular architecture we need to
capture such resource footprints of software modules at different
levels of granularity (e.g., program statements, blocks in procedure bodies
and whole procedures), and be able to compare different task
compositions. Consequently, the modeling language must feature
massively parallel operators over such task-level resource footprints~\cite{pbb08}.
A similar notion of resource footprints and composition can be used to
express the properties of the architecture in a machine model to
capture the resources that the architecture can make available to the
code.

Working with resource footprints can be supported by an abstract
behavioral specification language~\cite{johnsen10fmco}, in which
models describe both tasks and deployments.
These models can be used to predict the non-functional behavior of 
code before it is deployed, and to compare deployments using formal
methods. This requires a formal semantics for the specification
language that can be used to devise static analysis techniques.

When developing code from scratch using a model-based approach, the
resource footprints can be specified in tandem with the standard model
in a model-driven development~\cite{arki2013:757,pllanauml02,pllanaint08,fahringer04}.
However, when building such models from existing HPC code, monitoring
profiles of low- and medium-scale systems can be used to extract
resource footprints that approximate the resource consumption in terms
of probabilistic distributions.

\subsection{Static analysis and optimization}
\label{subsec:static}

The application of formal methods to parallel programs for analyzing functional properties, such as safety and liveness, has a long tradition. For \emph{non-functional properties}, such as execution time and energy consumption, most performance analysis approaches use \emph{monitoring} and present statistical information to the user. These approaches are helpful to improve HPC application code, but they also have some shortcomings:
\begin{enumerate}
	\item Due to non-determinism, different program executions might lead to different observations. As a consequence, these methods are not able to provide reliable probabilistic information about average or worst-case execution times. 
	\item They are based on execution on a real platform, thus they cannot be used to predict performance on Exascale computers, which are not available yet. 
	\item These methods can be used to identify execution bottlenecks, but they cannot explain the reasons for these bottlenecks, and thus they do not offer any concrete support for code improvement.
\end{enumerate}

We expect that \emph{formal methods} can address these limitations to provide performance analysis tools that considerably go beyond the state-of-the-art. A major step in this direction will be the usage of resource footprints which describe both HPC applications and execution platforms as abstract probabilistic models. Formal analyses can be applied to these models to predict their probabilistic behavior. While a range of techniques are available for non-probabilistic programs, the analysis of parallel probabilistic programs still need development effort.
To achieve a reasonable balance between scalability and precision for challenging HPC applications, it seems fundamental to use \emph{hybrid approaches}~\cite{pllanahyb09} that combine techniques such as static analysis, dynamic analysis, simulation, (parametric) model checking~\cite{jansen*:parametric}, counterexample-guided abstraction refinement~\cite{dehnert*:prism}, deductive approaches, etc.

To deliver the envisaged performance analysis tools, we face the following challenges: (1) determining the computation of cost properties that are given by means of probabilistic distributions; (2) the inference of average cost in addition to the traditional worst-case cost; (3) take into account the underlying platform through a set of probabilistic parameters; (4) deal with massive and heterogeneous parallelism~\cite{benkner11,kessler12,sandrieser12,dokulil12} which is challenging for program analysis in general; and (5) develop multi-objective resource usage analyses and optimizations.

\subsection{Runtime analysis and optimization}
\label{subsec:runtime}

Formal modeling and static analysis should be enhanced with analysis of measurements at runtime. Plenty of tools (for instance, \url{http://www.vi-hps.org}) have been developed for performance measurement and analysis of HPC applications at runtime. However, these tools will experience several issues when applied to Exascale. The collection rates and the overall volume of monitoring data in an Exascale computing environment will exceed the scalability of present performance tools. Therefore, throttling the data volume will have to be applied online in order to store as less data as possible and as much as necessary for later \emph{post mortem} analysis. However, simple profiling will not be sufficient due to loss of temporal information, thus a hybrid approach will have to be applied that performs on-the-fly trace analysis in order to discard irrelevant data, while retaining the same amount of information. 

The metric classification should be based on the formal model (see Section~\ref{subsec:modeling}). Such an approach will provide a generic insight into the performance of an HPC application that can be used for detecting performance bottlenecks. The instrumentation and hardware counter monitoring should follow a similar procedure where source code probing should be applied automatically by using tools such as OPARI \cite{opari}. While many tools for collecting metrics of computing performance have been developed, very few analysis tools exist for energy consumption metrics in adequate accuracy and time resolution necessary for the runtime performance analysis \cite{alonso_tools_2012,bohra_vmeter:_2010}. 

Currently, common approaches (see for example PRACE best practices \cite{prace-bp}) for optimizing HPC applications require per-case inspection of runtime performance measurement data, such as profiling and tracing data. After the critical region has been determined, diverse heuristic approaches, such as ``\textit{trial and error}'', ``\textit{educated guess}'' or ``\textit{rule of thumb}'', are applied to make changes in the affected source code sections. The most significant limitation of these heuristics and knowledge-based approaches is that,
\begin{enumerate}
	\item all changes are made directly and manually in the source code, and
	\item the effect of the changes does not always lead to an improved performance which makes necessary the repeating of all steps several times. 
\end{enumerate}

Moreover, Exascale computers pose a multi-objective optimization problem, weighing out the effects of several sometimes incongruent requirements. Therefore, a systematic and automatic approach for the optimization problem is essential to find the optimal solution. Another problem is that the critical section in an application typically changes with the optimization iterations and/or with upscaling, due to the law of diminishing returns, which makes the manual analysis and source code changes even more laborious and inefficient, even if done by an experienced HPC developer. Thus instrumentation, collection/measurement and analysis steps should be automated, for example based on high-level scalable tools \cite{scalasca,kojak,expert}, and integrated into a feedback loop (see Section~\ref{subsec:autonomic}).

\subsection{Autonomic computing}
\label{subsec:autonomic}

During the execution of an application, failures may occur or the application performance may be below the expectation. These issues are addressed typically by programmers in a ``\textit{trial and error}'' manner, i.e.\ by manually changing and adapting their code to handle the failures and improve the performance. Our proposed framework (Section \ref{sec:approach}) provides means for model-based failure handling or performance improvement based on \emph{autonomic computing}. Autonomic computing addresses self-managing characteristics of distributed computing resources with the facilities to adapting to unpredictable changes while hiding management complexity to operators and users \cite{kephart}. Among the explored categories, \emph{advanced-control based methods} and more specifically  \emph{distributed controllers} are the first candidates to realize autonomic computing in Exascale systems. 

We propose to devise methodologies to efficiently collect runtime
information balancing the amount and cost for storage of monitoring data with the quality of monitored data
necessary to make deductions about the application behavior (e.g.\ trace analysis). The goal is thereby to define
methodologies to scale current monitoring tools to Exascale, balancing between quality and volume of monitoring data. 

\begin{figure*}[ht]
\centering
\includegraphics[width=0.72\textwidth]{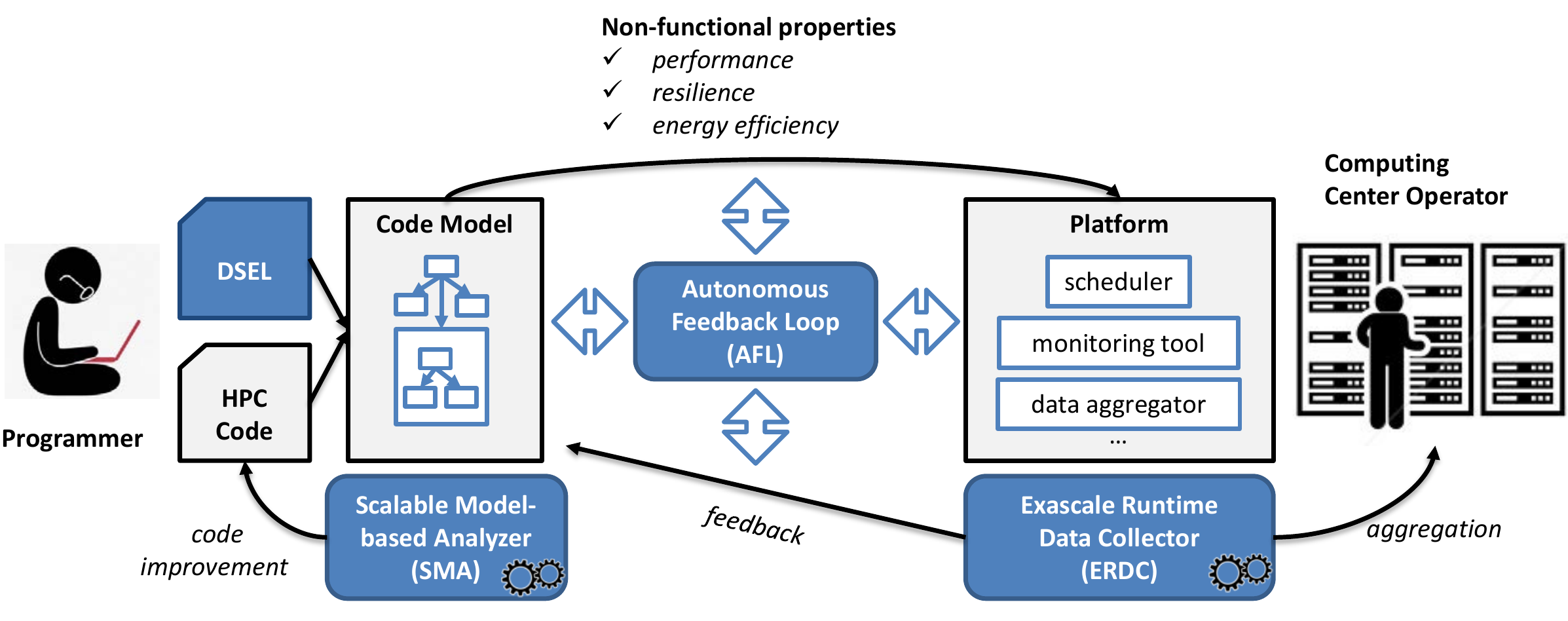}
\caption{Our conceptual framework for porting HPC applications to Exascale computing systems}
\label{fig:diagram}
\end{figure*}

Combining the information from both static code analysis and runtime analysis, as outlined above, we will iteratively apply objective-oriented transformations to legacy application code at a formal level based on the Exascale DSL model (see Section~\ref{subsec:modeling}). To this end, we will automate the analysis, optimization and transformation processes by implementing a generic feedback loop independent of the concrete programming language, algorithms used and target hardware architecture. A feedback loop driver enables to link the static analysis tool, the runtime analysis tool, the knowledge database and multi-parameter multi-objective optimization. As output, a set of rules (policies) is generated which is then
applied to transform the formal application model and to adapt the runtime environment parameters (cf.~Figure~\ref{fig:diagram}). After the transformations a new application executable is built and started in the adapted runtime
environment. This described loop is iterated until convergence of the optimization.

\section{Conceptual Framework and Benefits}
\label{sec:ab}

In this section we propose a conceptual framework for porting HPC applications to Exascale computing systems. Furthermore, we highlight benefits of our conceptual framework in the context of Exascale computing.

\subsection{Conceptual Framework}
\label{sec:approach}

Our proposed approach for preparing HPC applications for Exascale is depicted in Figure~\ref{fig:diagram}. The usage of a \emph{Domain-Specific Exascale Language (DSEL)} facilitates the programmer to express non-functional aspects (like required time to solution, resilience or energy-efficiency) of the execution of scalable parallel HPC codes. DSEL has a formal operational semantics that enables the formal analysis of the code. The aim of the \emph{Scalable Model-based Analyzer (SMA)} is to address non-functional properties of HPC codes, with a particular focus on scalability while complying with the crucial dimensions of resource consumption for Exascale computing: time, energy, and resilience. The SMA is responsible for analyzing resource consumption in terms of time, energy, and resilience, based on developed DSELs. The \emph{Exascale Runtime Data Collector (ERDC)} is responsible for scalable monitoring to \emph{extract important monitoring data} through the utilization of various techniques like \emph{filtering}, \emph{streaming}, or \emph{data mining}. The runtime information is used to verify or to tune the model of the code via the \emph{Autonomous Feedback Loop (AFL)}. To endow the system with self-adaption, control-theoretical concepts are incorporated in autonomic computing paradigm. Based on the autonomic technology for application optimization, programmers will be less dependent on the currently used ``\emph{trial-and-error}'' approach.

Our approach considers optimization opportunities during the application life cycle comprising improvements based on static code analysis, deployment-time optimization, and run-time optimization. The developed models are used to identify the potential for improvement of the scalability for HPC applications under study and suggest application modifications that may lead to better scalability.

\subsection{Benefits}
\label{sec:benefits}

Exascale computing is not simply the continuation of a computational capability trend that has been proven true for the last five decades. 
First, while clock rate scaling is limited, complex multicore architectures and parallel computing still follow Moore's law.
Second, Exascale computing capability will finally allow complex real-life simulations and data analytics. The latter will greatly expand the horizons of scientific discovery and enable the new data-driven economy to become a reality.

However, the Exascale promise faces a series of obstacles, with the most difficult being energy, scalability, reliability and programmability.  Our proposal is to develop a holistic, unifying and mathematically founded framework to systematically attack the roots of these problems. That is, instead of attacking these problems separately, we propose a holistic approach to study them as a multi-parametric problem which will allow us to deeply understand their interplay and thus make the right decisions to navigate in this complex landscape.

The benefits are targeting the full spectrum of actors and beneficiaries. System developers will have a much better path to design, while end users and application developers will benefit from increased scalability, performance, reliability and programmability.  HPC centers will see a great increase in overall system usability and an energy budget that is affordable. This in turn has the potential to greatly limit and contain the overall impact of high end HPC to the environment.

\section{Related Work}
\label{sec:related-work}

In a prospective analysis of issues with extreme scale systems \cite{sark2009hs:012045}, the importance of concurrency, energy efficiency and resilience of software, as well as software--hardware co-design has been elucidated. 

Focusing on energy-aware HPC numerical applications, the EXA2GREEN project (\url{http://exa2green-project.eu}) has developed energy-aware performance metrics \cite{beka2010c:187}, as well as energy-aware basic algorithm motifs such as linear solvers \cite{klav2014mb:20130278}. Further work will strongly benefit from these results. Different power measurement interfaces available on current architecture generations have been evaluated and the role of the sampling rate has been discussed \cite{Diouri201468}. 

The AutoTune approach \cite{mice2013cs:328} employs the Periscope tuning framework \cite{bene2010pg:1} to automate performance analysis and tuning of HPC applications with the goal to improve performance and energy efficiency. Therein, both performance analysis and tuning are performed automatically during a single run of the application. 

The DEEP project \cite{eick2013lm:885} has developed a novel Exascale-enabling supercomputing architecture with a matching software stack and a set of optimized grand-challenge simulation applications. 
The goal of the DEEP architecture is to enable unprecedented scalability and with an extrapolation to millions of cores to take the DEEP concept to an Exascale level. The follow-up DEEPer project
(\url{http://www.deep-er.eu}) is mainly focusing on I/O and resiliency aspects. 

The CRESTA project (\url{http://www.cresta-project.eu}) has adopted a co-design strategy for Exascale, including all aspects of hardware architectures, system and application software.
A major asset from the CRESTA project is the Score-P measurement system \cite{agui2013dg} on which  an integration and automation of performance analysis tools (cf.~Section~\ref{subsec:runtime}) can be based. In addition, efforts have been made on developing a domain-specific language for expressing parallel auto-tuning specifications and an adaptive runtime support framework. 

\section{Summary}
\label{sec:summary}

Exascale computing will revolutionize high-performance computing, but the first Exascale systems are not expected to appear before 2020. In this paper we have argued that the effort for preparing HPC applications for Exascale should start now. We have proposed that porting of HPC applications should be made by successive, stepwise improvements based on the currently available assumptions and data about Exascale systems. This approach should support application improvement each time new information about future Exascale systems becomes available, including the time when the application is actually deployed and runs on a concrete Exascale system. We have identified challenges that need to be addressed and recommended solutions in key areas of interest for our approach including: formal modeling, static analysis and optimization, runtime analysis and optimization, and autonomic computing. Our future research will address the development of a framework that supports the conceptual framework presented in this paper.



\end{document}